\documentclass[english,10pt, onecolumn, notitlepage,  superscriptaddress]{revtex4-1}
\usepackage[T1]{fontenc}
\usepackage[latin9]{inputenc}
\usepackage{amstext}
\usepackage{graphicx}

\def\7{$\;$}
\def\l{\left}
\def\r{\right}
\def\be{\begin{equation}}
\def\ee{\end{equation}}
\def\bea{\begin{eqnarray}}
\def\eea{\end{eqnarray}}
\def\f{\frac}

\def\no{\nonumber}
\def\d{{\rm d}}
\def\no{\nonumber}

\usepackage{babel}

\makeatother

\usepackage{babel}
\begin{document}

\title{Ricci curvature non-minimal derivative coupling cosmology with field re-scaling}

\author{Burin Gumjudpai}\email{buring@nu.ac.th}
\affiliation{The Institute for Fundamental Study ``The Tah Poe Academia Institute'', Naresuan University, Phitsanulok 65000, Thailand}
\author{Yuttana Jawralee}\email{yuttanaj57@email.nu.ac.th}
\affiliation{The Institute for Fundamental Study ``The Tah Poe Academia Institute'', Naresuan University, Phitsanulok 65000, Thailand}
\author{Narakorn Kaewkhao}\email{narakornk56@email.nu.ac.th}
\affiliation{The Institute for Fundamental Study ``The Tah Poe Academia Institute'', Naresuan University, Phitsanulok 65000, Thailand}

\date{\today}
\begin{abstract}
In this letter, cosmology of a simple NMDC gravity with  $\xi  R \phi_{,\mu}\phi^{,\mu}$ term and a free kinetic term is considered in flat geometry and in presence of dust matter. A logarithm field transformation $\phi' = \mu \ln \phi$ is proposed phenomenologically.  Assuming slow-roll approximation, equation of motion, scalar field solution and  potential are derived as function of kinematic variables. The field solution and potential are found straightforwardly for power-law, de-Sitter and super-acceleration expansions. Slow-roll parameters and slow-roll condition are found to depend on more than one variable. At large field the re-scaling effect can enhance the acceleration. For slow-rolling field, the negative coupling $\xi$ could enhance the effect of acceleration.

\end{abstract}
\maketitle

\section{Introduction}
\label{Section:Introduction}
Various observations such as supernova type Ia (SNIa)
 \cite{Amanullah2010, Astier:2005qq, Goldhaber:2001a, Perlmutter:1997zf, Perlmutter:1999a, Riess:1998cb,
 Riess:1999ar, RiessGold2004, Riess:2007a, Tonry:2003a},
large-scale structure surveys \cite{Scranton:2003, Tegmark:2004}, cosmic microwave background (CMB) anisotropies
 \cite{Larson:2010gs, arXiv:1001.4538, CMBXRay:2014, Masi:2002hp} and X-ray luminosity from galaxy clusters \cite{CMBXRay:2014, Allen:2004cd, Rapetti:2005a} agree on recent cosmic
acceleration. Yet to be known form of dark energy is believed to drive the acceleration
  \cite{Copeland:2006a, Padmanabhan:2004av, Padmanabhan:2006a, AT:DE2010}. As far as the
combined CMB data is concerned, the data favors the universe with a cosmological constant and a dark matter, that
is the $\Lambda$CDM model. Severe inconsistency in the value of the observed value of $\Lambda$ and the prediction of the quantum
field theory motivates attempt to find alternative theory to cope with the present small value of dark energy. Amongst
these attempts are dynamical scalar field models, for examples, quintessence
 \cite{Caldwell:1997ii} and k-essence \cite{Chiba:1999ka, ArmendarizPicon:2000dh, ArmendarizPicon:2000ah} and modified gravity such as braneworlds, $f(R)$, scalar-tensor theories  (see e.g.  \cite{Nojiri:2010wj, AntoShijiLRR2010, Carroll2004} for review). When the scalar field is
non-minimally coupling to the gravity part, the acceleration is possible. Coupling function $f(\phi, \phi_{,\mu}, \phi_{,\mu\nu}, \ldots)$ can be motivated
in scalar quantum electrodynamics or in gravitational theory of which Newton's constant is a function of the density \cite{Amendola1993}. Capozziello \cite{Capozziello:1999xt} found that other possible coupling terms apart from
 $R \phi_{,\mu}\phi^{,\mu}$
and  $R^{\mu\nu} \phi_{,\mu} \phi_{,\nu}$ are not necessary. These terms can be lower energy limits of higher dimensional
theories or Weyl anomaly of $\mathcal{N} = 4$ conformal supergravity
\cite{Liu:1998bu, Nojiri:1998dh}. The scalar derivative term coupling to the gravitational sector is known as non-minimal derivative coupling-NMDC and
 theory with such term, e.g.  $\kappa_1 R \phi_{,\mu}\phi^{,\mu}$  and
 $\kappa_2 R^{\mu\nu} \phi_{,\mu} \phi_{,\nu}$  without  $V(\phi)$ nor $\Lambda$ could give de Sitter expansion  \cite{Capozziello:1999uwa}
 and this fact attracts cosmological consideration. NMDC models with these coupling terms were studied  with further modifications \cite{Granda:2010hb, Granda:2010ex, Granda:2011zk}. A special case of when there are two terms with two coupling constants related by $\kappa \equiv   \kappa_2  =  -2 \kappa_1$ gives rise to a theory of two NMDC terms making the Einstein tensor, $G^{\mu\nu}$ coupled to the scalar field derivative term as seen in literatures \cite{Sushkov:2009, Saridakis:2010mf, Gao:2010vr, Germani:2010gm, Skugoreva:2013ooa, Sushkov:2012, Koutsoumbas:2013boa, Darabi:2013caa, Germani:2009, Sadjadi:2012zp, Tsujikawa:2012mk, Ema:2015oaa, Jinno:2013fka, Ema:2016hlw, Yang:2015pga, Sadjadi:2010bz, Gumjudpai:2016ioy}. The NMDC term is indeed a special case of the Horndeski action \cite{Horn}.
  Recent review of the studies in this area can be seen in Ref. \cite{Gumjudpai:2015vio}.

 In this letter, we consider the simplest NMDC model with  $\xi  R \phi_{,\mu}\phi^{,\mu}$ as the only NMDC coupling term and a free kinetic term. We propose modification to the model to investigate the effect of NMDC term and effect of the field transformation to the acceleration with motivation that the superpotential at early time could evolve to dark energy potential at late time. We aim to check if the model could be a phenomenologically candidate of inflation and dark energy.
We derive cosmological field equation of this model in a flat FLRW universe.  We give a correction to a previous similar result. Assuming power-law, de Sitter and super-acceleration expansions, the scalar  field solutions and scalar potential functions are found. The slow-roll parameters and acceleration condition are derived with consideration of a superpotential.

\section{Cosmology of the NMDC gravity coupling to Ricci curvature}
In this work, we consider the action in Jordan frame,
  \bea     \label{S1}
 S &=& \int {\d}^4x\sqrt{-g} \bigg[\frac{1}{16\pi G}R - \frac{1}{2}\partial_{\mu}\phi' \partial^{\mu}\phi' - \frac{1}{2}\xi R\partial_{\mu}\phi' \partial^{\mu}\phi'
  - V(\phi') \bigg]  + S_{\rm m}.
\eea
  The Lagrangian density has dimension of ${mass}^4$, the scalar field $\phi'$ has dimension of $mass$ and the Ricci curvature has the dimension of ${length}^{-2} = {mass}^2 $. A transformation $\phi '   \rightarrow  \phi' =  \mu \ln{\phi} $ is proposed here phenomenologically so that  the field $\phi$ is dimensionless. 
  It has been shown that Horndeski action is invariant under rescaling transformation like $\phi' \rightarrow \phi = s(\phi') \phi'$ \cite{Bettoni:2013diz} which is a motivation of our rescaling.
  The mass dimension of the field prior to the transformation ($\phi'$) is of the constant $\mu$.
 Result of the transformation is an action with re-scaling field in all terms of the  field function,
    \bea
 S &=& \int {\d}^4x\sqrt{-g} \bigg[\frac{1}{16\pi G}R - \frac{1}{2} \l(\f{\mu^2}{\phi^2}\r) \partial_{\mu}\phi \partial^{\mu}\phi - \frac{1}{2}\xi R  \l(\f{\mu^2}{\phi^2}\r) \partial_{\mu}\phi \partial^{\mu}\phi
  - V(\phi) \bigg] + S_{\rm m}.
   \label{eq_feil}
\eea
All terms need to have dimension of ${mass}^4$. Hence, in this theory, the coupling constant $\xi$ has dimension of $mass^{-2}$ or ${length}^2$.
Previously  the simplest $\xi  R \phi_{,\mu}\phi^{,\mu}$ NMDC model was studied and the scalar potential was derived assuming power-law expansion \cite{GrandaColumbia}. The model with free kinetic term and the two NMDC terms:
$\kappa_1 R \phi_{,\mu}\phi^{,\mu}$,
 $\kappa_2 R^{\mu\nu} \phi_{,\mu} \phi_{,\nu}$ modulating with field re-scaling factors $1/\phi^2$  was phenomenologically proposed also by Granda in Ref. \cite{Granda:2009fh}. Having the re-scaling factor in his model, the coupling constant, $ \kappa_1$ and  $\kappa_2$ become dimensionless. When both of the NMDC terms are dominant, they play the role of dark matter, giving dust expansion solution. In order to obtain present acceleration, the potential is needed in the theory. In our model, we have only one NMDC term, $\xi  R \phi_{,\mu}\phi^{,\mu}$ and the modulation with field re-scaling  $1/\phi^2$ is achieved by field transformation.
As $T_{\mu\nu}  = T_{\mu \nu}^{(\phi)} + T_{\mu \nu}^{(\rm m)} $,  varying the scalar part of the action (\ref{eq_feil}) in the metric formalism, we obtain
   \bea
T_{\mu \nu}^{(\phi)} &=& \f{\mu^2}{\phi^2}\nabla_{\mu}\phi \nabla_{\nu}\phi -\f{1}{2}g_{\mu\nu} \l(\f{\mu^2}{\phi^2} \nabla_{\rho}\phi \nabla^{\rho}\phi \r) - g_{\mu\nu}V(\phi) \no \\
       & & + \xi \Bigg[ \l(R_{\mu\nu} - \f{1}{2} g_{\mu\nu}R \r) \l(\f{\mu^2}{\phi^2}\nabla_{\rho}\phi \nabla^{\rho}\phi \r) +  R \l(\f{\mu^2}{\phi^2}\nabla_{\mu}\phi \nabla_{\nu}\phi \r) \no \\
       & & -  g_{\mu\nu} \nabla_{\rho} \nabla^{\rho} \l(\f{\mu^2}{\phi^2} \nabla_{\sigma}\phi \nabla^{\sigma}\phi \r) +  \nabla_{\mu} \nabla_{\nu} \l(\f{\mu^2}{\phi^2} \nabla_{\rho}\phi \nabla^{\rho}\phi\r) \Bigg].
\eea
Hence the Einstein field equation is 
\be
\l(1  -  8 \pi G  \xi \f{\mu^2}{\phi^2} \nabla_{\rho}\phi  \nabla^{\rho}\phi    \r)G_{\mu\nu}   \;=\; 8 \pi G \l[  T^{(\phi, \rm eff)}_{\mu \nu}   +   T_{\mu \nu}^{(\rm m)} \r]
\ee
where $T^{(\phi, \rm eff)}_{\mu \nu}   \equiv T_{\mu \nu}^{(\phi)} -  \xi G_{\mu\nu} \l(\f{\mu^2}{\phi^2} \nabla_{\rho}\phi  \nabla^{\rho}\phi  \r) $. 
The effective gravitational constant is hence
\be
G_{\rm eff}   =  \f{G}{\l(1  -  8 \pi G  \xi \f{\mu^2}{\phi^2} \nabla_{\rho}\phi  \nabla^{\rho}\phi    \r)}\,.
\ee
In this work, we do not modify geometry term ($G_{\mu\nu}$) but let the NMDC modification be included in the energy-momentum tensor term, i.e. in the pressure and density so that we use standard gravitational constant in the evolution equations.  We consider flat FLRW universe with $
 {\rm d}s^2  = -{\rm d}t^2 + a^2(t){\rm d}\textbf{x}^{2}\,
$ in this work.
The (00) and ($ii$) components of tensor $T_{\mu\nu}^{(\phi)}$, are as follows
\bea
\rho_{\rm \phi} = T_{00}^{(\phi)} = \f{\mu^2}{2}\f{\dot{\phi}^2 }{\phi^2}+ V(\phi)+ 3\mu^2 \xi \bigg[(2\dot{H} + 3H^2)\f{\dot{\phi}^2}{\phi^2} + 2H\f{\dot{\phi}}{\phi^2}\ddot{\phi} - 2H\f{\dot{\phi^3}}{\phi^3}\bigg],
\eea
and
\bea
 P_{\rm \phi} = \f{T_{11}^{(\phi)}}{a^2} = \f{\mu^2}{2}\f{\dot{\phi}^2 }{\phi^2} - V(\phi) +
\mu^2 \xi \bigg[(2\dot{H}+3H^2)\f{\dot{\phi^2}}{\phi^2} - 4H\f{\dot{\phi}}{\phi^2}\ddot{\phi} + 4H\f{\dot{\phi^3}}{\phi^3}  \no \\
+2\bigg(\f{\ddot{\phi^2}}{\phi^2} + \f{\dot{\phi}}{\phi^2}\dddot{\phi} - 5\f{\dot{\phi^2}}{\phi^3}\ddot{\phi} + 3\f{\dot{\phi^4}}{\phi^4}\bigg)\bigg].
\eea
 Our results give minor correction to the result given in Ref. \cite{GrandaColumbia}.  Having dust matter action, $S_{\rm m}$,
 the $(00)$ and $(ii)$ components of the Einstein field equation gives
\be
H^2 = \f{8 \pi G}{3}(\rho_{\phi} + \rho_{\rm m}), \;\;\;\;
2\dot{H}+ 3H^2  =  - 8 \pi G (P_{\phi} + P_{\rm m}).
\ee
Combining these two field equations, therefore
\be
\dot{H} =  - 4 \pi G (P_{\rm tot} + \rho_{\rm tot}),    \label{eqdotH}
\ee
where ${P_{\rm m}}=0, P_{\phi}, {\rm \rho_{\rm m}}$ and ${\rho_{\phi}}$  are the pressure and density of dust and scalar field respectively. Therefore $P_{\rm tot} = {P_{\rm m}} +  P_{\phi} $ and   $\rho_{\rm tot}  = {\rm \rho_{\rm m}}+{\rho_{\phi}}$.
 Varying the action (\ref{S1}) with respect to another dynamical variable $\phi'$ to obtain Euler-Lagrange equation of the NMDC field,
and transforming $\phi'$ to $\phi$,  hence
\bea
\ddot{\phi}+3H\dot{\phi} -\f{\dot{\phi}^{2}}{\phi}-\f{\dot{\phi}^{2}}{\phi}\xi \l(6\dot{H}+12H^2  \r)+ 6\xi\ddot{H}\dot{\phi} + 6\xi H  \l(7\dot{H}+6H^2 \r)  \dot{\phi}
+\xi \l(6\dot{H}+12H^2 \r)\ddot{\phi} + \f{\phi^{2}}{\mu^2}\f{\rm d }{\rm d\phi}V(\phi) = 0.   \label{eq_kgg}
\eea

\section{Slow-roll approximation}
Under slow-roll assumption of $0\ll |\dot{\phi}|\ll 1$ and $|\dddot{\phi}|\ll |\ddot{\phi}|\ll |\dot{\phi}|$ we neglect the terms with $\dot{\phi}^3, \ddot{\phi}^2, \dot{\phi}\dddot{\phi}, \dot{\phi}^2\ddot{\phi}$ and $\dot{\phi}^4$ hence the pressure and density are
\bea
\rho_{\rm \phi}  \; \simeq  \;  \f{\mu^2}{2}\f{\dot{\phi}^2 }{\phi^2}+ V(\phi)+ 3\mu^2 \xi \l[(2\dot{H} + 3H^2)\f{\dot{\phi}^2}{\phi^2} + 2H\f{\dot{\phi}}{\phi^2}\ddot{\phi}   \r],
\label{eq_rhoapprox}
\eea
and
\bea
 P_{\rm \phi} \; \simeq  \;  \f{\mu^2}{2}\f{\dot{\phi}^2 }{\phi^2} - V(\phi) +
\mu^2 \xi \l[(2\dot{H}+3H^2)\f{\dot{\phi^2}}{\phi^2} - 4H\f{\dot{\phi}}{\phi^2}\ddot{\phi} \r]. \label{eq_Papprox}
\eea
Using these approximated quantities, the Eq.  (\ref{eqdotH}) is
\bea
\dot{H} &  \simeq & -\f{ 8 \pi G }{2} \l[ \mu^{2}\f{\dot{\phi}^2}{\phi^2} + 4 \mu^{2} \xi\l(2\dot{H}+3H^2\r) \f{\dot{\phi}^2}{\phi^2} + 2 \mu^{2} \xi H \f{\dot{\phi}}{\phi^2} \ddot{\phi} +\rho_{\rm m} \r]\,.  \label{eqdhotHlong}
\eea
Consider linear approximation of the the NMDC field equation,
\bea
\ddot{\phi} \approx -3H\dot{\phi} - \bigg(\f{\phi^{2}}{\mu^{2}}\bigg)\f{\rm d}{\rm d\phi}V(\phi).   \label{eq_line}
\eea
Hence Eq. (\ref{eqdhotHlong}) is approximated as
\bea
\dot{H}
& \simeq & - 4 \pi G \l[      \l( 1 +    8\xi \dot{H} +  6\xi H^{2}     \r)
\mu^{2}\f{\dot{\phi}^2}{\phi^2} - 2\xi H \dot{\phi}\f{\rm d}{\rm d\phi}V(\phi) + \rho_{\rm m}  \r],  \label{eq_Hdotfromfield}
\eea
which recovers the standard GR form when $\xi \rightarrow 0$, that is
$\dot{H} =  - 4 \pi G \l(       \dot{\phi}'^2     + \rho_{\rm m}  \r).$ Eq. (\ref{eq_Hdotfromfield}) is a combination of two field equations.
There is  another way of expressing  $\dot{H}$ by considering time derivative of the Friedmann equation,
\bea
 \dot{H} = \f{4 \pi G}{3 H}   \l(\dot{\rho}_{\rm \phi}+\dot{\rho}_{\rm m}\r)\,.   \label{eq_FRderi}
\eea
The term $\dot{\rho}_{\phi}$ can be found from Eq. (\ref{eq_rhoapprox}). Using approximation (\ref{eq_line}),
\bea
\dot{\rho}_{\phi}  &\simeq& 3\l[
 \f{\dot{\phi}^{2}}{\phi^{2}} \mu^2  \l(
 - H
 - 12 \xi H\dot{H}
 + 2 \xi\ddot{H}
 - 18 \xi H^{3} \r)
  -6\xi \dot{\phi}  \f{\rm d}{\rm d\phi}V(\phi)    \l(  \dot{H}         + H^{2}   \r)           \r],
\eea
which reduces to GR case, $\dot{\rho}_{\phi} = - 3H \dot{\phi}'^2 $, as $\xi \rightarrow 0$. Using this relation and $\dot{\rho}_{\rm m} = -3 H \rho_{\rm m}$, hence
\bea
\dot{H} &\simeq&   - 4 \pi G \l[  \l(1  + 12 \xi \dot{H}  - 2 \xi \f{\ddot{H}}{H}  + 18 \xi H^2  \r) \f{\dot{\phi}^2}{\phi^2} \mu^2  \,+ \, 6 \xi \dot{\phi} \f{\d V}{\d \phi} \l(\f{\dot{H}}{H} + H   \r)   + \rho_{\rm m}                                                \r]
\eea
or rewritten as
\bea
\f{\d V}{\d \phi}  & \simeq & \f{- H \dot{H}     -   4 \pi G \l(  H  + 12 \xi \dot{H}H - 2 \xi \ddot{H}  + 18 \xi H^3   \r) \l( \f{\dot{\phi}^2}{\phi^2} \mu^2 \r)  -  4 \pi G \rho_{\rm m} H   }{24 \pi G \xi \dot{\phi}  \l( \dot{H} + H^2    \r) }
\eea
This expression of $\d V/\d \phi$ comes from the time derivative of the Friedmann equation (\ref{eq_FRderi}). We use it in the other $\dot{H}$ equation (\ref{eq_Hdotfromfield}) so that
\be
\mu^2 \f{\dot{\phi}^2}{\phi^2}  \;\simeq\;  \f{-6 \dot{H}^2 - 8 H^2 \dot{H}    -   8 \pi G \rho_{\rm m} \l(3 \dot{H}  + 4 H^2\r)    }{
 8\pi G    \l[   3 \dot{H}   + 4 H^2 + \xi \l(36 H^4   + 54 H^2 \dot{H}     + 24 \dot{H}^2  - 2 H \ddot{H}     \r)       \r] }\,.   \label{eq_erty}
\ee
Since $\rho_{\rm m} = \rho_{\rm m, 0} a^{-3} $, hence Eq. (\ref{eq_erty}) can be integrated to give a scalar field solution,
\be
\phi(t)  \;\simeq\; \exp \l\{\f{1}{\mu} \int \d t \l[\f{-6 \dot{H}^2 - 8 H^2 \dot{H}    - 8 \pi G \rho_{\rm m}(a) \l(3 \dot{H}  + 4 H^2\r)    }{
 8\pi G    \l[   3 \dot{H}   + 4 H^2 + \xi \l(36 H^4   + 54 H^2 \dot{H}     + 24 \dot{H}^2  - 2 H \ddot{H}     \r)       \r] }\r]^{1/2}\r\}\,.
\ee
Using the Friedmann equation, the scalar field density (\ref{eq_rhoapprox}), the linear  approximation in Eq. (\ref{eq_line}) and Eq.  (\ref{eq_Hdotfromfield}), the scalar potential is derived,
\bea
V(a, H,\dot{H},\ddot{H}) &\;  \simeq \; & \f{6\dot{H}}{8\pi G} + \f{3H^2}{8\pi G} + 4\rho_{\rm m}(a) + \f{1}{2}\l(5+54\xi H^2+36\xi\dot{H}\r) \no \\
& & \times\l[\f{-6 \dot{H}^2 - 8 H^2 \dot{H}    - 8 \pi G \rho_{\rm m}(a) \l(3 \dot{H}  + 4 H^2\r)    }{
 8\pi G    \l[   3 \dot{H}   + 4 H^2 + \xi \l(36 H^4   + 54 H^2 \dot{H}     + 24 \dot{H}^2  - 2 H \ddot{H}     \r)       \r] }\r].
\eea
If exact form of $a = a(t)$ is known, one can straightforwardly derive the scalar field solution and the scalar potential.
\section{Case study: explicitly known forms of expansion}
The cosmic system in the late universe where the ingredients are dark energy and dark matter could be viewed as the presence of the NMDC field with dust matter fluid. In this section we assume the known form of expansion so that  the scalar field solution and the scalar potential can be derived straightforwardly. The expansions considered are of power-law, de-Sitter and super-acceleration types.
\subsection{Power-law expansion}
We consider an expansion function, $a \propto t^{p}$ where $p>0$. To find the exact scalar field solution of this case, we need to neglect small contribution of $\rho_{\rm m}$ in order to obtain exact solution,
\bea \label{Fieldoftime}
\phi(t) \simeq \phi_{0}\exp \l[\l(\f{\sqrt{2p}}{\mu \sqrt{8\pi G}}\r) {\rm arcsinh}\l(\f{t}{\sqrt{p \xi}\, \alpha}\r)\r]\,,
\eea
where $\alpha^2   =  \f{36p^2-54p+20}{4p-3}  \;  \simeq \;  9p - \f{27}{4}    $
 and $\chi = {\mu \sqrt{8\pi G}}/{\sqrt{2p}} \;$.
The scalar potential is found,
\bea
V(\phi) &\simeq&        \f{3(p-2)}{8 \pi G \xi   \alpha^2 \sinh^2[ \chi\ln({\phi}/{\phi_{0}})]} \,+\,
 \f{5  \alpha^2 \sinh^2 [\chi \ln({\phi}/{\phi_{0}})] + 54 p - 36 }{ {8 \pi G \xi   \alpha^4 \sinh^2[ \chi\ln({\phi}/{\phi_{0}})]}       \cosh^2 [\chi \ln({\phi}/{\phi_{0}})]
      }  .
\eea
\subsection{de-Sitter expansion}
The de-Sitter expansion, $a \propto e^{H_{0}t}$ is assumed and we keep the dust matter contribution in the solution here.  The scalar solution is
\bea
\phi(t)  \:&\simeq&\:   \phi_{0} \exp\l[\f{2}{3\mu H_0}\sqrt{\f{\rho_{\rm m,0}\exp(-3H_{0}t)}{ (1 + 9\xi H_{0}^{2})}} \, \r]\,,
\eea
with  the potential\,
\bea
V(\phi)
&\simeq & \f{3H_{0}^{2}}{8\pi G} +  \f{27}{8} H_0^2 \mu^2 \l(  1 + 6 \xi H_0^2 \r)      \l[ \ln \l(\f{\phi}{\phi_{0}}\r)\r]^2.
\eea

\subsection{Super-acceleration expansion}
We assume super-acceleration expansion, $a = a_0 \l[(t_{\rm s}- t)/(t_{\rm s} - t_0) \r]^q$ where $q < 0$ and $t_{\rm s}$ is the future singularity. Neglecting dust matter density, one can find the solution
 \bea
\phi(t) = \phi_{0}\exp \l[-\l(\f{\sqrt{2 |q|}}{\mu \sqrt{8\pi G}}\r) {\rm arcsin}\l(\f{t_{\rm s}-t}{ \sqrt{|q| \xi}\,   \beta}\r)\r],
\eea
where $\beta^2 = \f{36q^2-54q+20}{4q-3}   \simeq \;  9q - \f{27}{4}$ and $\chi = {\mu \sqrt{8\pi G}}/{ \sqrt{2|q|}} \;$. Hence
\bea
V(\phi) &=&  \f{3(q-2)}{- 8\pi G \xi \beta^2  \sin^2[ \chi\ln({\phi}/{\phi_{0}})]}\,+\,
 \f{- 5  \beta^2 \sin^2 [ \chi \ln({\phi}/{\phi_{0}})] + 54 q - 36 }{ {- 8 \pi G \xi   \beta^4 \sin^2[ \chi\ln({\phi}/{\phi_{0}})]}       \cos^2 [ \chi \ln({\phi}/{\phi_{0}})]
     },
\eea
where we have used relations, $\sinh (-i x) = - i \sin (x)$ and $\arcsin(-x) = i {\rm arcsinh} (i x) $.

\section{Slow-roll parameters and acceleration condition}
Considering inflationary epoch, the scalar field is dominant.  Eq. (\ref{eq_kgg}) can be approximated further as in the slow-roll regime, $\dot{\phi}^2\simeq 0,\, \ddot{\phi}\simeq 0$ and  $|\ddot{H}|\ll |H\dot{H}| \ll |H^{3}|$.
\bea
\dot{\phi} \simeq
-  \f{\phi^{2} V_{\phi} }{\mu^2 3 H (1 + 12 \xi H^2)}   \label{eq_kggA}
\eea
where $V_{\phi} \equiv {\rm d }V/{\rm d\phi} $. This equation reduces to GR case, $\dot{\phi}' \simeq -  {V_{\phi'} }/{(3 H)}$ for $\xi = 0$. The $\dot{H}$ equation is approximated as the GR case with potential dominated,
\be
\dot{H}\,  \simeq \,  \f{V_{\phi}  \dot{\phi}}{6HM^2_{\rm P}}\,  \simeq \,  \f{\sqrt{3} V_{\phi} \dot{\phi}}{6 \sqrt{V} M_{\rm P}}   
 \label{eq_fufu}
\ee
to which we will use $\dot{\phi}$ of Eq. (\ref{eq_kggA}). We set here $M_{\rm P}^2 \equiv 8 \pi G$. Hence
\be
\dot{H}  \simeq   -\f{\phi^2 V_{\phi}^2}{18 \mu^2 H^2 M_{\rm P}^2  (1 + 12 \xi H^2)}
\ee
and the slow-roll parameter,
\be
\epsilon_{\rm v}  \equiv   - \f{\dot{H}}{H^2}  \simeq     \f{M_{\rm P}^2}{2}    \l(\frac{V_{\phi}}{V}\r)^2  \f{\phi^2}{\mu^2 (1 + 12 \xi\,{H^2})}.
\ee
The factor $\phi^2/\mu^2$ is an effect of the field re-scaling and the $
\xi$ term is the NMDC effect.   The slow-roll parameter $\delta \equiv \ddot{\phi}/(H \dot{\phi})$ is
\be
\delta   =  \epsilon_{\rm v} \l[ -\f{4 V}{\phi V_{\phi}}   +   \f{(1 + 12 \xi V/M_{\rm P}^2)}{(1 + 4 \xi V/M_{\rm P}^2)} \r]   -   \eta_{\rm v}
\ee
where
\be
\eta_{\rm v}  =  M_{\rm P}^2 \l( \f{V_{\phi \phi}}{V}\r)   \f{\phi^2}{\mu^2 (1 + 12 \xi\,{H^2})}.
\ee
The re-scaling effect makes the
slow-roll condition $|\delta| \ll 1$ depending on counterbalancing of potential $V$ and $V_{\phi}$ terms in the square bracket. It also depends on the factor ${\phi^2}/{\mu^2 (1 + 12 \xi\,{H^2})}$ of $\eta_{\rm v}$  and $\epsilon_{\rm v}$.
As a result, the spectral index $n_{\rm s} = 1 - 4\epsilon_{\rm v}  - 2 \delta$ also depends on these factors.  GR limit of $\delta$ can not be achieved by setting $\phi = \mu = 1$ and $\xi =  0$  because derivative of $\phi^2 V_{\phi}$ in (\ref{eq_kggA})  renders the $-4V/(\phi V_{\phi})$ term which does not vanish in the GR limit.  The acceleration condition is found as
\be
\dot{\phi}^2   \; < \;    \f{(\phi^2/\mu^2) V(\phi)}{\l[   1  + \xi  \l( \f{\sqrt{3} V_{\phi} \dot{\phi}}{\sqrt{V} M_{\rm P}}   +   \f{3 V(\phi)}{M_{\rm P}^2}     \r)  \r]}\,.
\ee
At large field, the factor  $(\phi^2/\mu^2)$  makes it easier for the  acceleration condition to be satisfied. If the potential has positive slope, the  $\xi < 0$ case makes the right hand side larger hence helping the condition to be satisfied easier. On the other hand, if $\xi > 0$ the right hand side is less, making it harder to achieve the acceleration.  Let us consider specific form of potential which is still a function of $\phi'$.  The exponential potential in supergravity model is \cite{Brax1, Co1}
\be
V(\phi') \: = \: \mathcal{M}^4 e^{- \lambda  \sigma \phi'}\,.  \label{eq_esi}
\ee
The transformation $\phi '   \rightarrow  \phi' =  \mu \ln{\phi} $ gives superpotential
 \be
V(\phi) \: = \: {M}^{4 + \lambda  \sigma \mu} \phi^{- \lambda  \sigma \mu}\,,
\ee
where $\mathcal{M}^4 \equiv {M}^{4 + \lambda  \sigma \mu} $. It has been known that there exists scaling solution of the scalar field under the potential (\ref{eq_esi})  \cite{Co1}. The scalar field may evolve under the superpotential at early time and afterwards at late time evolving as dark energy under the exponential function (\ref{eq_esi}) with scaling solution.
If we consider a simple case of $\lambda  \sigma \mu =  2$, the acceleration condition reads
\be
\dot{\phi}^2 \: < \:   \f{(M^6/\mu^2) \phi^2 }{\phi^2 + \xi \l( -\sqrt{12} \f{M^3}{M_{\rm P}}   \dot{\phi}   +  3 \f{M^6}{M_{\rm P}^2}           \r)}\,.
\ee
The right hand side can be considered as effective potential. However, the potential here has negative slope hence the role of sign of $\xi$ in enhancing the acceleration needs to be reconsidered. If the field speed is in the range $\dot{\phi} < \sqrt{3} M^3/(2 M_{\rm P})$, the $\xi < 0$ is able to enhance the acceleration.

\section*{Conclusions}
  In this letter, we have derived cosmological field equations of a proposed phenomenological NMDC model containing $\xi  R \phi_{,\mu}\phi^{,\mu}$ term with a logarithm field transformation $\phi' = \mu \ln \phi$ in a flat universe as motivated by  invariance of Horndeski action \cite{Bettoni:2013diz}. This work modifies the action as well as gives minor corrections of the results given in Refs. \cite{GrandaColumbia} and \cite{Granda:2009fh}. Assuming slow-roll approximation, we derive scalar field equations of motion. In presence of dust matter, scalar field solution and scalar potential are derived as function of kinematic variables $a,H, \dot{H}$ and  $\ddot{H}$. Therefore, the field solution, $\phi(t)$ and the potential, $V=V(\phi)$ can be found straightforwardly for explicit forms of the expansion function $a(t)$. Dynamically significant expansion functions: power-law, de Sitter and super-acceleration expansions are assumed and scalar field solutions and potential are derived.  We found that the slow-roll parameters and slow-roll condition depend on more than one condition and can not reduce to the GR case. The acceleration condition is modified and at large field the re-scaling effect can enhance the acceleration. For slow-rolling field, the negative coupling $\xi$ could also enhance the acceleration.

\section*{Acknowledgments}
This work is supported by the National Research Council of Thailand (R2558B132).

\end{document}